\documentclass{article}
\usepackage{amssymb}
\usepackage{amsmath}

\input{tcilatex}
\usepackage[T1]{fontenc}
\begin{document}

\title{The Gravitational Field of a Radiating Electromagnetic Dipole}
\author{T.M. Adamo$^{1}$ \& E.T. Newman$^{2}$ \\
$^{1}$Dept of Mathematics, $^{2}$Dept of Physics and Astronomy\\
\textit{University of Pittsburgh, Pgh., PA 15260, USA}\\
$^{1}$tma8@pitt.edu, $^{2}$newman@pitt.edu}
\maketitle

\begin{abstract}
We begin with the time-dependent electric and magnetic dipole solution of
Maxwell's equations in Minkowski space. This Maxwell field is then used to
determine the behavior of the gravitational field (the Weyl tensor) as a
second-order perturbation off of the Minkowski background. From the Weyl
tensor we go on and find the spin-coefficients and the full metric in this
approximation. The physical meaning of many of the relations is discussed.
In particular we can identify the conservation law of angular momentum that
contains an angular momentum flux term.
\end{abstract}

\section{Introduction}

Recently Bramson\cite{1} has investigated, perturbatively, the gravitational
effects due to a radiating pure electric dipole Maxwell field as source. \
He began with the pure Maxwell field in a Minkowski background space-time.
He then integrated the Bianchi Identities (also on the Minkowski background)
for the linearized Weyl tensor with the given Maxwell field inserted into
the stress tensor acting as the source for the Weyl tensor. He found, as
expected, that indeed the Maxwell field did \textquotedblleft
stimulate\textquotedblright\ a gravitational response including energy loss
and gravitational radiation. He stopped short here and made no attempt to
find the perturbed metric. \ All of Bramson's work, as well as the present
work, is done in the spin-coefficient formalism

The purpose of the present note is three-fold: 1. we give a mild
generalization of Bramson's work by including both electric \textit{and}
magnetic dipoles in the calculations, 2. \ we correct a minor error in
Bramson's calculation, and 3. (our major contribution), we continue the
perturbation calculation to include the integration of the perturbed
spin-coefficients and metric variables, thus obtaining the perturbed metric.

The strategy is to integrate the Maxwell equations and the Bianchi
Identities (as did Bramson) and then use these results in the integration of
the complete set of spin-coefficient and metric equations. The metric is
then easily constructed.

In Section 2 we write the time dependent electric and magnetic dipole
solution of the Maxwell field, assuming no Coulomb charge. In Section 3, we
integrate the Bianchi identities in order to obtain the behavior of the Weyl
tensor (i.e., the gravitational field). In Section 4 the spin coefficients
and metric variables are found. As will be seen, all the gravitational
information will be coded into the Bondi shear, which is found as a function
of the electromagnetic dipole field. In Section 5, a reality condition (an
algebraic Weyl tensor symmetry) is applied to the known Weyl components.
This leads to physical meaning for many of the Weyl components as well as an
understanding of the gravitational dynamics. In Section 6 the space-time
metric is easily calculated and given. Conclusions and comments are given in
Section 7.

\section{The Maxwell Field}

In flat Minkowski space, we choose the well-known Bondi coordinate system $%
(u,r,\zeta ,\bar{\zeta}),$ i.e., the coordinate system where $u$ labels the
light cones with apex on a time-like world-line, $\zeta =e^{i\phi }\cot
(\theta /2),$ the complex stereographic coordinates labeling the individual
null geodesics, while $r$\ is the affine parameter along the null geodesics.
The Bondi null tetrad $\{l^{a},n^{a},m^{a},\bar{m}^{a}\}$ is such that $%
l^{a} $ is tangent to the null geodesics. This, with a further choice of the
tetrad, fixes the flat space spin-coefficients (SC) as:

\begin{eqnarray}
\kappa &=&\sigma =\varepsilon =\pi =\tau =\lambda =\gamma =\nu =0,  \label{a}
\\
&&%
\begin{array}{lll}
\rho =-r^{-1}; & \alpha =\ \frac{-\zeta }{2r}=\ \frac{\alpha ^{0}}{r}, & 
\beta =\ \frac{\bar{\zeta}}{2r}=\ \frac{\beta ^{0}}{r},%
\end{array}
\notag \\
&&%
\begin{array}{l}
\mu =\ \frac{\mu ^{0}}{r}=\ \frac{-1}{r}.%
\end{array}
\notag
\end{eqnarray}

In this tetrad and coordinate system, the retarded flat-space Maxwell field
for electric and magnetic dipole radiation can be written as\cite{Max.,N.T.}:

\begin{subequations}
\begin{eqnarray}
\phi _{0} &=&\ \frac{\phi _{0}^{0}}{r^{3}}\equiv \ \frac{2D^{i}}{r^{3}}%
Y_{1i}^{1},  \label{b} \\
\phi _{1} &=&\ \frac{\phi _{1}^{0}}{r^{2}}+\ \frac{\phi _{1}^{1}}{r^{3}}%
\equiv \frac{\sqrt{2}D^{i\prime }}{r^{2}}Y_{1i}^{0}-\ \frac{D^{i}}{r^{3}}%
Y_{1i}^{0},  \notag \\
\phi _{2} &=&\ \frac{-2D^{i\prime \prime }}{r}Y_{1i}^{-1}+\ \frac{2\sqrt{2}%
D^{i\prime }}{r^{2}}Y_{1i}^{-1}-\ \frac{D^{i}}{2r^{3}}Y_{1i}^{-1},  \notag \\
\phi _{2}^{0} &\equiv &-2D^{i\prime \prime }Y_{1i}^{-1}.  \notag
\end{eqnarray}

The \textit{complex} three-vector $D^{i}$ (the complex dipole moment) can be
written as 
\end{subequations}
\begin{equation}
D^{i}=D_{E}^{i}+iD_{M}^{i},  \label{c}
\end{equation}%
where $D_{E}^{i}$ and $D_{M}^{i}$ are the electric and magnetic dipole
moments respectively. $D^{i}$ is an arbitrary function of the retarded time, 
$u_{r}\equiv \sqrt{2}u.$

For our perturbation calculations we will treat, $D^{i}$ (or $\xi ^{i}$) as
first order.

\textbf{Remark}: Often\cite{3,4,5} we find it useful to write $D^{i}\equiv
q\xi ^{i},$ where $\xi ^{i}$ is treated as a complex position vector.

\textbf{Remark:} In the preceding as well as in what follows, we will denote
the derivative with respect to the Bondi time as $\partial _{u}()=(^{\cdot
}) $, while differentiation with respect to the retarded Bondi time is $%
\partial _{u_{r}}()=(^{\prime })$. Note that $(^{\cdot })=\sqrt{2}(^{\prime
})$. \ Later in Section 5, just to view the equations with appropriate units
(i.e., to restore the $c$), we make the transition $(^{\prime
})=>c^{-1}(^{\prime })$ in a few equations.\qquad \qquad

\section{The Gravitational Field-the Weyl Tensor}

We now turn to the linear Bianchi identities in the SC formalism with the
flat space set of coefficients given in (1). The radial equations can be
written as\cite{N.T.}:

\begin{equation}
\ \frac{\partial \psi _{1}}{\partial r}+\ \frac{4\psi _{1}}{r}=\frac{\bar{%
\eth }\psi _{0}}{r}+\frac{5k\phi _{0}^{0}\bar{\phi}_{1}^{0}}{r^{6}}+\ \frac{%
k\eth (\phi _{0}^{0}\bar{\phi _{0}^{0}})-4k\phi _{0}^{0}\eth \bar{\phi}%
_{0}^{0}}{r^{7}},  \label{1}
\end{equation}%
$\ $ 
\begin{eqnarray*}
\frac{\partial \psi _{2}}{\partial r} &=&-\ \frac{3\psi _{2}}{r}\ +\frac{%
\bar{\eth }\psi _{1}}{r}-\frac{2k\phi _{1}^{0}\bar{\phi}_{1}^{0}}{r^{5}}+%
\frac{k}{3r^{6}}{\Large (}4\phi _{1}^{0}\eth \bar{\phi}_{0}^{0}+4\bar{\phi}%
_{1}^{0}\bar{\eth }\phi _{0}^{0}-2\eth (\phi _{1}^{0}\bar{\phi}_{0}^{0})+%
\bar{\eth }(\phi _{0}^{0}\bar{\phi}_{1}^{0}){\Large )} \\
&&-\ \ \frac{k}{3r^{7}}[\frac{5}{2}\bar{\eth }\phi _{0}^{0}\eth \bar{\phi}%
_{0}^{0}-\eth (\bar{\phi}_{0}^{0}\bar{\eth }\phi _{0}^{0})+\frac{1}{2}\bar{%
\eth }(\phi _{0}^{0}\eth \bar{\phi}_{0}^{0})+\phi _{0}^{0}\bar{\phi}%
_{0}^{0}]-k\Delta (\frac{\phi _{0}^{0}\bar{\phi}_{0}^{0}}{3r^{6}}),
\end{eqnarray*}

\begin{eqnarray*}
\ \frac{\partial \psi _{3}}{\partial r} &=&-\ \frac{2\psi _{3}}{r}\ +\frac{%
\bar{\eth }\psi _{2}}{r}-\frac{k\phi _{2}^{0}\bar{\phi}_{1}^{0}}{r^{4}}+%
\frac{k}{3r^{5}}{\Large (}2\phi _{2}^{0}\eth \bar{\phi}_{0}^{0}+4\bar{\phi}%
_{1}^{0}\bar{\eth }\phi _{1}^{0}-\eth (\phi _{2}^{0}\bar{\phi}_{0}^{0})+2%
\bar{\eth }(\phi _{1}^{0}\bar{\phi}_{1}^{0}){\Large )} \\
&&-\ \ \frac{k}{3r^{6}}{\Large (}\frac{5}{2}\eth \bar{\phi}_{0}^{0}\bar{\eth 
}\phi _{1}^{0}+\frac{5}{4}\bar{\phi}_{1}^{0}\bar{\eth }^{2}\phi
_{0}^{0}-\eth (\bar{\phi}_{0}^{0}\bar{\eth }\phi _{1}^{0})+\bar{\eth }(\phi
_{1}^{0}\eth \bar{\phi}_{0}^{0})+\bar{\eth }(\bar{\phi}_{1}^{0}\bar{\eth }%
\phi _{0}^{0}){\Large )} \\
&&+\ \ \frac{k}{12r^{7}}{\Large (}2\bar{\eth }(\bar{\eth }\phi _{0}^{0}\eth 
\bar{\phi}_{0}^{0})-\eth (\bar{\phi}_{0}^{0}\bar{\eth }^{2}\phi
_{0}^{0})+3\eth \bar{\phi}_{0}^{0}\bar{\eth }^{2}\phi _{0}^{0}{\Large )}-%
\frac{2k}{3}\Delta (\frac{\phi _{1}^{0}\bar{\phi}_{0}^{0}}{r^{5}}-\frac{\bar{%
\phi}_{0}^{0}\bar{\eth }\phi _{0}^{0}}{2r^{6}}){\Large ,}
\end{eqnarray*}

\begin{eqnarray*}
\ \frac{\partial \psi _{4}}{\partial r} &=&-\ \frac{\psi _{4}}{r}\ +\frac{%
\bar{\eth }\psi _{3}}{r}+\frac{\bar{\eth }(\phi _{2}^{0}\bar{\phi}_{1}^{0})}{%
r^{4}}-\frac{k}{2r^{5}}{\Large (}\bar{\eth }(\phi _{2}^{0}\eth \bar{\phi}%
_{0}^{0})+2\bar{\eth }(\bar{\phi}_{1}^{0}\bar{\eth }\phi _{1}^{0})-2\phi
_{2}^{0}\bar{\phi}_{0}^{0}{\Large )} \\
&&+\frac{k}{4r^{6}}{\Large (}2\bar{\eth }(\bar{\eth }\phi _{1}^{0}\eth \bar{%
\phi}_{0}^{0})+\bar{\eth }(\bar{\phi}_{1}^{0}\bar{\eth }^{2}\phi _{0}^{0})-4%
\bar{\phi}_{0}^{0}\bar{\eth }\phi _{1}^{0}{\Large )}-\frac{k(\bar{\eth }%
(\eth \bar{\phi}_{0}^{0}\bar{\eth }^{2}\phi _{0}^{0})-2\bar{\phi}_{0}^{0}%
\bar{\eth }^{2}\phi _{0}^{0})}{8r^{7}} \\
&&-k\Delta {\Large (}\frac{\phi _{2}^{0}\bar{\phi}_{0}^{0}}{r^{4}}-\frac{%
\bar{\phi}_{0}^{0}\bar{\eth }\phi _{1}^{0}}{r^{5}}+\frac{\bar{\phi}_{0}^{0}%
\bar{\eth }^{2}\phi _{0}^{0}}{4r^{6}}{\Large ).}
\end{eqnarray*}

In these equations, the flat space operator $\Delta $ is given by: $\Delta
\equiv \partial _{u}-\partial _{r}$ and the constant $k=2G/c^{4}$ . If Eqs.(%
\ref{b}) had been used here a large simplification would have occurred.

Now, as we are interested only in those gravitational effects which arise
due to the Maxwell field, Eq.(\ref{b}), we can follow Bramson\cite{1} in
setting $\psi _{0}=0$. In this case, integrating Eqs.(\ref{1}), with
frequent use of Clebsch-Gordon expansions\cite{spin-s} to re-express the
quadratic terms, produces: 
\begin{eqnarray}
\psi _{1} &=&\ \frac{\psi _{1}^{0}}{r^{4}}-ik{\Large (}\frac{10D^{i}\bar{D}%
^{j\prime }}{r^{5}}-\frac{3\sqrt{2}D^{i}\bar{D}^{j}}{r^{6}}{\Large )}%
\epsilon _{ijk}Y_{1k}^{1}  \label{3} \\
&&-\ \ k{\Large (}\frac{5\sqrt{2}D^{i}\bar{D}^{j\prime }}{r^{5}}-\frac{3D^{i}%
\bar{D}^{j}}{r^{6}}{\Large )}Y_{2ij}^{1},  \notag
\end{eqnarray}

\begin{eqnarray}
\psi _{2} &=&\ \frac{\psi _{2}^{0}}{r^{3}}-\frac{\bar{\eth }\psi _{1}^{0}}{%
r^{4}}+k{\Large (}\frac{2D^{i\prime }\bar{D}^{j\prime }}{r^{4}}-\frac{(2+8%
\sqrt{2})(\bar{D}^{i}D^{j})^{\prime }}{9r^{5}}+\frac{20D^{i}\bar{D}^{j}}{%
9r^{6}}{\Large )}\delta _{ij}  \label{4} \\
&&+ik{\Large (}\frac{24D^{i}\bar{D}^{j\prime }+(4+\sqrt{2})(\bar{D}%
^{i}D^{j})^{\prime }}{r^{5}}-\frac{\sqrt{2}D^{i}\bar{D}^{j}}{r^{6}}{\Large )}%
\epsilon _{ijk}Y_{1k}^{0}  \notag \\
&&+k{\Large (}\frac{4D^{i\prime }\bar{D}^{j\prime }}{3r^{4}}+\frac{30\sqrt{2}%
D^{i}\bar{D}^{j\prime }+(1+2\sqrt{2})(\bar{D}^{i}D^{j})^{\prime }}{18r^{5}}-%
\frac{10D^{i}\bar{D}^{j}}{9r^{6}}{\Large )}Y_{2ij}^{0},  \notag
\end{eqnarray}

\begin{eqnarray}
\psi _{3} &=&\frac{\psi _{3}^{0}}{r^{2}}-\ \ \frac{\bar{\eth }\psi _{2}^{0}}{%
r^{3}}+\ \ \frac{\bar{\eth }^{2}\psi _{1}^{0}}{2r^{4}}  \label{5} \\
&&+ki{\Large (}\frac{2D^{i\prime \prime }\bar{D}^{j\prime }}{r^{3}}-\frac{2(%
\sqrt{2}D^{i\prime }\bar{D}^{j\prime }+(1-\sqrt{2})(\bar{D}^{i}D^{j\prime
})^{\prime })}{3r^{4}}  \notag \\
&&+\frac{(34-\sqrt{2})(D^{i}\bar{D}^{j})^{\prime }-D^{i}\bar{D}^{j\prime }}{%
9r^{5}}-\frac{5\sqrt{2}D^{i}\bar{D}^{j}}{4r^{6}}{\Large )}\epsilon
_{ijk}Y_{1k}^{-1}  \notag \\
&&-k{\Large (}\frac{\sqrt{2}D^{i\prime \prime }\bar{D}^{j\prime }}{r^{3}}-%
\frac{20D^{i\prime }\bar{D}^{j\prime }+\sqrt{2}(\bar{D}^{i}D^{j\prime
})^{\prime }}{3r^{4}}{\Large )}Y_{2ij}^{-1}  \notag \\
&&-k{\Large (}\frac{37\sqrt{2}D^{i\prime }\bar{D}^{j}+(3-49\sqrt{2})(D^{i}%
\bar{D}^{j})^{\prime }}{18r^{5}}-\frac{23D^{i}\bar{D}^{j}}{12r^{6}}{\Large )}%
Y_{2ij}^{-1},  \notag
\end{eqnarray}

\begin{eqnarray}
\psi _{4} &=&\frac{\psi _{4}^{0}}{r}-\frac{\bar{\eth }\psi _{3}^{0}}{r^{2}}+%
\frac{\bar{\eth }^{2}\psi _{2}^{0}}{2r^{3}}-\frac{\bar{\eth }^{3}\psi
_{1}^{0}}{6r^{4}}  \label{6} \\
&&-k{\Large (}\frac{3\sqrt{2}D^{i\prime \prime }\bar{D}^{j\prime
}+2(D^{i\prime \prime }\bar{D}^{j})^{\prime }}{r^{3}}-\frac{(16\sqrt{2}-72)(%
\bar{D}^{i}D^{j\prime })^{\prime }}{9r^{4}}  \notag \\
&&-\frac{128D^{i\prime }\bar{D}^{j\prime }}{9r^{4}}+\frac{(24-107\sqrt{2}%
)(D^{i}\bar{D}^{j})^{\prime }-25\sqrt{2}D^{i\prime }\bar{D}^{j}}{36r^{5}}+%
\frac{47D^{i}\bar{D}^{j}}{15r^{6}}{\Large )}Y_{2ij}^{-2}.  \notag
\end{eqnarray}

The radially independent `functions of integration' $\psi _{1}^{0}$, $\psi
_{2}^{0}$, $\psi _{3}^{0}$, and $\psi _{4}^{0}$ are now determined by the
non-radial Bianchi identities. These take the form\cite{N.T.}:

\begin{equation}
\eth \psi _{1}^{0}=3k\phi _{0}^{0}\bar{\phi}_{2}^{0},  \label{7}
\end{equation}

\begin{equation}
\dot{\psi}_{1}^{0}=-\eth \psi _{2}^{0}+2k\phi _{1}^{0}\bar{\phi}_{2}^{0},
\label{8}
\end{equation}

\begin{equation}
\dot{\psi}_{2}^{0}=-\eth \psi _{3}^{0}+k\phi _{2}^{0}\bar{\phi}_{2}^{0},
\label{9}
\end{equation}

\begin{equation}
\dot{\psi}_{3}^{0}=-\eth \psi _{4}^{0}.  \label{10}
\end{equation}

A straightforward `integration' of (\ref{7}), i.e., just the comparison of
spherical harmonic coefficients, yields:

\begin{equation}
\psi _{1}^{0}=3kD^{i}\bar{D}^{j\prime \prime }Y_{2ij}^{1}+a^{k}Y_{1k}^{1}.
\label{11}
\end{equation}

Here, the radially independent complex 3-vector $a^{k}(u_{r})$ emerges as
another `function of integration', the behavior of which will be determined
shortly. Continuing on to (\ref{8}), we can again , by comparing spherical
harmonic coefficients and the use of a Clebsch-Gordon expansion of the
Maxwell field term, obtain:

\begin{equation}
\psi _{2}^{0}=\chi +{\Large (}\frac{\sqrt{2}}{2}a^{k\prime }-2kiD^{i\prime }%
\bar{D}^{j\prime \prime }\epsilon _{ijk}{\Large )}Y_{1k}^{0}+\sqrt{2}k%
{\Large (}\frac{1}{2}(D^{i}\bar{D}^{j\prime \prime })^{\prime }+\frac{1}{3}%
D^{i\prime }\bar{D}^{j\prime \prime }{\Large )}Y_{2ij}^{0}.  \label{12}
\end{equation}

The scalar valued $\chi (u_{r})$ enters here in the same manner as the $%
a^{k} $ entered in the previous

\noindent integration. Now, from (\ref{9}), noting that since $\psi _{3}$
has spin-wt $s=-1\ $and hence \textit{it cannot have}

\noindent an $l=0$ harmonic contribution, leaves us with the following
relations:

\begin{eqnarray}
\psi _{3}^{0} &=&{\Large [}2\sqrt{2}ki(D^{i\prime }\bar{D}^{j\prime \prime
})^{\prime }\epsilon _{ijk}+\sqrt{2}ki\bar{D}^{i\prime \prime }D^{j\prime
\prime }\epsilon _{ijk}-a^{k\prime \prime }{\Large ]}Y_{1k}^{-1}  \label{13}
\\
&&+\ k{\Large [}\frac{1}{3}\bar{D}^{i\prime \prime }D^{j\prime \prime
}-(D^{i}\bar{D}^{j\prime \prime })^{\prime \prime }-\ \frac{2}{3}(D^{i\prime
}\bar{D}^{j\prime \prime })^{\prime }{\Large ]}Y_{2ij}^{-1},  \notag
\end{eqnarray}

\begin{equation}
\dot{\chi}=\sqrt{2}\chi ^{\prime }=\ \frac{4k}{3}\bar{D}^{i\prime \prime
}D^{j\prime \prime }\delta _{ij}.  \label{14}
\end{equation}

It should be noted that the evolution equation for $\chi $ is equivalent to
that given by Bramson \cite{1}. From a physical viewpoint it is (up to
multiplicative constants) exactly the Bondi energy loss theorem. It
coincides exactly with the classical E\&M electric and magnetic dipole
energy loss.

Finally from Eq.(\ref{10}), recalling that $\psi _{4}$ has spin weight $%
s=-2, $ it follows that there \textit{cannot} be an $l=1$ harmonic
contribution in that equation. This allows us to determine $a^{k}$ (up to
initial values) as well as obtain the desired expression for $\psi _{4}^{0}$:

\begin{equation}
\psi _{4}^{0}=\sqrt{2}k\left[ (D^{i}\bar{D}^{j\prime \prime })^{\prime
\prime \prime }-\ \frac{1}{3}(\bar{D}^{i\prime \prime }D^{j\prime \prime
})^{\prime }+\frac{2}{3}(D^{i\prime }\bar{D}^{j\prime \prime })^{\prime
\prime }\right] Y_{2ij}^{-2},  \label{15}
\end{equation}

\begin{equation}
a^{k\prime \prime \prime }=\sqrt{2}ki\{2(D^{i\prime }\bar{D}^{j\prime \prime
})^{\prime \prime }+(\bar{D}^{i\prime \prime }D^{j\prime \prime })^{\prime
}\}\epsilon _{ijk}.  \label{C2}
\end{equation}

\textbf{Remark: }The Weyl tensor components $\psi _{1}$, $\psi _{2}$, and $%
\psi _{3}$ all appear to contain the $l=1$ spherical harmonic contributions
from the vector $a^{k}$; however

\noindent Eq.(\ref{10}) puts a condition, Eq.(\ref{C2}), on the $a^{k}$ that
eliminates the $l=1$ part of $\psi _{3}^{0}$. Bramson$\cite{1},$improperly
set the $a^{k}=0,$ apparently overlooking this \noindent

\noindent condition.

\textbf{Remark: }There is a further equation (a reality condition) on $\psi
_{2}^{0}$ that plays a very important role. It however involves a further
variable (the spin-coefficient $\sigma ^{0}$, \ i.e., \ the Bondi shear) for
its description. We postpone a discussion of this condition until later.

\section{The Spin Coefficients}

In order to determine the metric of the perturbed space-time with the Weyl
tensor given in Section 3, we need the full set of spin coefficients. We
integrate the spin-coefficient equations on a flat background but `driven'
by the just obtained Weyl tensor. \ Afterwards we integrate the `metric'
equations which then lead to the metric expression.

Since we considered the Maxwell field to be small (first order) and since
they entered the Bianchi Identities quadratically we must consider the Weyl
tensor to be second order. We thus will calculate the spin-coefficients with
a second order correction to the zeroth order flat space coefficients.

The only SCs which explicitly remain zero in the perturbation are ($\kappa $%
, $\pi $,$\epsilon )$ due to our choice of the tetrad\cite{7}. This tetrad
choice also continues to fix $\rho =\bar{\rho}$ and $\tau =\bar{\alpha}%
+\beta $.

We first integrate the \textquotedblleft optical
equations\textquotedblright\ 
\begin{equation}
\ \frac{\partial \rho }{\partial r}=\rho ^{2}+\sigma \bar{\sigma}+\frac{%
k\phi _{0}^{0}\bar{\phi}_{0}^{0}}{r^{6}},  \label{A}
\end{equation}

\begin{equation}
\ \frac{\partial {\sigma }}{\partial r}=2\rho \sigma .  \label{B}
\end{equation}%
for the spin coefficients $\rho $ and $\sigma $, the divergence and shear of 
$l^{a}.$Writing $\rho =r^{-1}+p$ and $\sigma =s$, we obtain: 
\begin{equation}
\rho =-\frac{1}{r}+\ \frac{p_{2}^{0}}{r^{2}}-\frac{\sigma ^{0}\bar{\sigma}%
^{0}}{r^{3}}-\frac{k\phi _{0}^{0}\bar{\phi}_{0}^{0}}{3r^{5}},  \label{C}
\end{equation}

\begin{equation}
\sigma =\ \frac{\sigma ^{0}}{r^{2}}.  \label{D}
\end{equation}

However, as the $\sigma ^{0}$ (our function of integration) is of second
order in the perturbation, the third term in the expression for $\rho $ can
be neglected. Furthermore, the integration constant $p_{2}^{0}$ can be set
equal to zero by a second order shift of the origin of the radial coordinate 
$r$, and our final expressions become:

\begin{equation}
\begin{array}{cc}
\rho =\ \frac{-1}{r}-\ \frac{k\phi _{0}^{0}\bar{\phi}_{0}^{0}}{3r^{5}}, & 
\sigma =\ \frac{\sigma ^{0}}{r^{2}}.%
\end{array}
\label{E}
\end{equation}

Solving the other radial and non-radial SC equations yields the following
results, which are in the leading terms equivalent to those of the
asymptotically flat space-time\cite{N.T.}:

\begin{equation}
\tau =-\frac{\psi _{1}^{0}}{2r^{3}}+ki{\Large (}\frac{8D^{i}\bar{D}^{j\prime
}}{3r^{4}}-\frac{\sqrt{2}D^{i}\bar{D}^{j}}{2r^{5}}{\Large )}\epsilon
_{ijk}Y_{1k}^{1}+k{\Large (}\frac{4\sqrt{2}D^{i}\bar{D}^{j\prime }}{3r^{4}}-%
\frac{D^{i}\bar{D}^{j}}{2r^{5}}{\Large )}Y_{2ij}^{1},  \label{F}
\end{equation}

\begin{eqnarray}
\alpha &=&\frac{\alpha ^{0}}{r}-\frac{\beta ^{0}\bar{\sigma}^{0}}{r^{2}}+ki%
{\Large (}\frac{2\bar{D}^{i}D^{j\prime }}{3r^{4}}-\frac{\sqrt{2}\bar{D}%
^{i}D^{j}}{4r^{5}}{\Large )}\epsilon _{ijk}Y_{1k}^{-1}  \label{G} \\
&&-\ k{\Large (}\frac{\sqrt{2}\bar{D}^{i}D^{j\prime }}{3r^{4}}-\frac{\bar{D}%
^{i}D^{j}}{4r^{5}}{\Large )}Y_{2ij}^{-1}+\frac{k\alpha ^{0}\phi _{0}^{0}\bar{%
\phi}_{0}^{0}}{12r^{5}},  \notag
\end{eqnarray}

\begin{eqnarray}
\beta &=&\ \frac{\beta ^{0}}{r}-\frac{\alpha ^{0}\sigma ^{0}}{r^{2}}-\frac{%
\psi _{1}^{0}}{2r^{3}}+ki{\Large (}\frac{10D^{i}\bar{D}^{j\prime }}{3r^{4}}-%
\frac{3\sqrt{2}D^{i}\bar{D}^{j}}{4r^{5}}{\Large )}\epsilon _{ijk}Y_{1k}^{1}
\label{H} \\
&&+k{\Large (}\frac{5\sqrt{2}D^{i}\bar{D}^{j\prime }}{3r^{4}}-\frac{3D^{i}%
\bar{D}^{j}}{4r^{5}}{\Large )}Y_{2ij}^{1}+\frac{k\beta ^{0}\phi _{0}^{0}\bar{%
\phi}_{0}^{0}}{12r^{5}},  \notag
\end{eqnarray}

\begin{eqnarray}
\gamma &=&-\ \frac{\psi _{2}^{0}}{2r^{2}}+\frac{\bar{\eth }\psi _{1}^{0}}{%
3r^{3}}+\frac{(\alpha ^{0}\psi _{1}^{0}+\beta ^{0}\bar{\psi}_{1}^{0})}{6r^{3}%
}  \label{I} \\
&&-k{\Large (}\frac{10D^{i\prime }\bar{D}^{j\prime }}{9r^{3}}-\frac{(1+7%
\sqrt{2})(D^{i}\bar{D}^{i})^{\prime }}{18r^{4}}+\frac{26D^{i}\bar{D}^{j}}{%
45r^{6}}{\Large )}\delta _{ij}  \notag \\
&&-ki{\Large (}\frac{24D^{i}\bar{D}^{j\prime }+(4+\sqrt{2})(D^{i}\bar{D}%
^{j})^{\prime }}{24r^{4}}-\frac{\sqrt{2}D^{i}\bar{D}^{j}}{6r^{5}}{\Large )}%
\epsilon _{ijk}Y_{1k}^{0}  \notag \\
&&-\ k{\Large (}\frac{2D^{i\prime }\bar{D}^{j\prime }}{3r^{3}}+\frac{(1-4%
\sqrt{2})(D^{i}\bar{D}^{j})^{\prime }+30\sqrt{2}D^{i}\bar{D}^{j\prime }}{%
24r^{4}}-\frac{7D^{i}\bar{D}^{j}}{45r^{5}}{\Large )}Y_{2ij}^{0}+\mathfrak{B},
\notag \\
\mathfrak{B} &\equiv &ik\beta ^{0}(\frac{8\bar{D}^{i}D^{j\prime }}{9r^{3}}-%
\frac{\sqrt{2}D^{j}\bar{D}^{i}}{16r^{4}})\epsilon _{ijk}Y_{1k}^{-1}-k\beta
^{0}(\frac{4\sqrt{2}\bar{D}^{i}D^{j\prime }}{9r^{3}}-\frac{D^{j}\bar{D}^{i}}{%
16r^{4}})Y_{2ij}^{-1}  \notag \\
&&-ik\alpha ^{0}(\frac{8D^{i}\bar{D}^{j\prime }}{9r^{3}}-\frac{\sqrt{2}\bar{D%
}^{j}D^{i}}{16r^{4}})\epsilon _{ijk}Y_{1k}^{1}-k\alpha ^{0}(\frac{4\sqrt{2}%
D^{i}\bar{D}^{j\prime }}{9r^{3}}-\frac{\bar{D}^{j}D^{i}}{16r^{4}})Y_{2ij}^{1}
\notag
\end{eqnarray}

\begin{equation}
\lambda =\frac{\lambda ^{0}}{r}+\frac{\bar{\sigma}^{0}}{r^{2}}+k{\Large (}%
\frac{2D^{i\prime \prime }\bar{D}^{j}}{r^{3}}-\frac{4\sqrt{2}D^{i\prime }%
\bar{D}^{j}}{3r^{4}}+\frac{D^{i}\bar{D}^{j}}{2r^{5}}{\Large )}Y_{2ij}^{-2},
\label{J}
\end{equation}

\begin{eqnarray}
\mu &=&{\small -}\frac{1}{r}{\small -}\frac{\psi _{2}^{0}}{r^{2}}+\frac{\bar{%
\eth }\psi _{1}^{0}}{2r^{3}}{\small -}k{\Large (}\frac{D^{i\prime }\bar{D}%
^{j\prime }}{r^{3}}{\small -}\frac{(2+8\sqrt{2})(\bar{D}^{i}D^{j})^{\prime }%
}{27r^{4}}+\frac{5D^{i}\bar{D}^{j}}{3r^{5}}{\Large )}\delta _{ij}  \label{Ja}
\\
&&-\ \ ki{\Large (}\frac{24D^{i}\bar{D}^{j\prime }+(4+\sqrt{2})(D^{i}\bar{D}%
^{j})^{\prime }}{18r^{4}}-\frac{\sqrt{2}D^{i}\bar{D}^{j}}{4r^{5}}{\Large )}%
\epsilon _{ijk}Y_{1k}^{0}  \notag \\
&&-\ k{\Large (}\frac{2D^{i\prime }\bar{D}^{j\prime }}{3r^{3}}+\frac{(1+2%
\sqrt{2})(D^{i}\bar{D}^{j})^{\prime }+30\sqrt{2}D^{i}\bar{D}^{j\prime }}{%
54r^{4}}-\frac{5D^{i}\bar{D}^{j}}{18r^{5}}{\Large )}Y_{2ij}^{0},  \notag
\end{eqnarray}

\begin{eqnarray}
\nu &=&-\ \frac{\psi _{3}^{0}}{r}+\frac{\bar{\eth }\psi _{2}^{0}}{2r^{2}}-%
\frac{(\bar{\psi}_{1}^{0}+\bar{\eth }^{2}\psi _{1}^{0})}{6r^{3}}  \label{K}
\\
&&-ki{\Large (}\frac{2D^{i\prime \prime }\bar{D}^{j\prime }}{r^{2}}-\frac{8%
\sqrt{2}D^{i\prime }\bar{D}^{j\prime }+\sqrt{2}D^{i\prime \prime }\bar{D}%
^{j}+(1-\sqrt{2})(\bar{D}^{i}D^{j\prime })^{\prime }}{9r^{3}}{\Large )}%
\epsilon _{ijk}Y_{1k}^{-1}  \notag \\
&&-\ ki{\Large (}\frac{8D^{i}\bar{D}^{j\prime }+6\bar{D}^{i}D^{j\prime }+(34-%
\sqrt{2})(D^{i}\bar{D}^{j})^{\prime }}{36r^{4}}-\frac{7\sqrt{2}\bar{D}%
^{i}D^{j}}{20r^{5}}{\Large )}\epsilon _{ijk}Y_{1k}^{-1}  \notag \\
&&+\ k{\Large (}\frac{\sqrt{2}D^{i\prime \prime }\bar{D}^{j\prime }}{r^{2}}-%
\frac{3D^{i\prime \prime }\bar{D}^{j}+26D^{i\prime }\bar{D}^{j\prime }+\sqrt{%
2}(\bar{D}^{i}D^{j\prime })^{\prime }}{9r^{3}}{\Large )}Y_{2ij}^{-1}  \notag
\\
&&+k{\Large (}\frac{70\sqrt{2}D^{i\prime }\bar{D}^{j}+(3-40\sqrt{2})(D^{i}%
\bar{D}^{j})^{\prime }}{72r^{4}}+\frac{11\bar{D}^{i}D^{j}}{60r^{5}}{\Large )}%
Y_{2ij}^{-1}.  \notag
\end{eqnarray}

The radially independent constants entering into the formulation are
governed by the following relations which emerge directly from the
non-radial SC equations:

\begin{eqnarray}
\alpha ^{0} &=&-\bar{\beta}^{0}=\frac{-\zeta }{2}\ ,  \label{L} \\
\ \lambda ^{0} &=&\sqrt{2}\overline{\sigma }^{0\prime },  \notag \\
\psi _{3}^{0} &=&\sqrt{2}\eth \overline{\sigma }^{0\prime },  \notag \\
\psi _{4}^{0} &=&-\sqrt{2}\lambda ^{0\prime }=-2\overline{\sigma }^{0\prime
\prime }.  \notag
\end{eqnarray}

First, however, by comparing the $\psi _{4}^{0}$ from Eqs.(\ref{15}) and (%
\ref{L}), i.e., 
\begin{equation}
\psi _{4}^{0}=-\sqrt{2}k{\Large [}\frac{1}{3}(\bar{D}^{i\prime \prime
}D^{j\prime \prime })^{\prime }-(D^{i}\bar{D}^{j\prime \prime })^{\prime
\prime \prime }-\ \frac{2}{3}(D^{i\prime }\bar{D}^{j\prime \prime })^{\prime
\prime }{\Large ]}Y_{2ij}^{-2},  \label{M}
\end{equation}%
with $\psi _{4}^{0}=-\dot{\lambda}^{0}=-\overline{\ddot{\sigma}}^{0},$ we
obtain the asymptotic Bondi shear 
\begin{equation}
\overline{\sigma }^{0}=\frac{\sqrt{2}}{2}k{\Large [}\frac{1}{3}\int (\bar{D}%
^{i\prime \prime }D^{j\prime \prime })du_{r}-(D^{i}\bar{D}^{j\prime \prime
})^{\prime }-\ \frac{2}{3}(D^{i\prime }\bar{D}^{j\prime \prime }){\Large ]}%
Y_{2ij}^{-2},  \label{N}
\end{equation}%
and the Bondi news function,%
\begin{equation}
\overline{\sigma }^{0\prime }=\frac{\sqrt{2}}{2}k{\Large [}\frac{1}{3}(\bar{D%
}^{i\prime \prime }D^{j\prime \prime })-(D^{i}\bar{D}^{j\prime \prime
})^{\prime \prime }-\ \frac{2}{3}(D^{i\prime }\bar{D}^{j\prime \prime
})^{\prime }{\Large ]}Y_{2ij}^{-2}.  \label{O}
\end{equation}

These are basic and important relations.

With these results \textit{all} the functions of integration are now known
in terms of the Maxwell field, or equivalently in terms of the complex $%
D^{i} $ and its time derivatives.

\section{The Reality Condition}

In the previous section we mentioned that there existed a further condition
on the asymptotic Weyl tensor, namely a \textit{reality condition}. By
defining the \textit{mass aspect, }$\Psi ,$ by

\begin{equation}
\Psi =\psi _{2}^{0\,}+\text{\dh }^{2}\overline{\sigma }^{0}+\sqrt{2}\sigma
^{0}(\overline{\sigma }^{0})^{\prime },  \label{aa}
\end{equation}%
the \textit{reality condition} is simply

\begin{equation}
\Psi =\overline{\Psi }.  \label{bb}
\end{equation}%
\qquad

Before we find the consequences of this condition it is worthwhile to see
the physical meaning of the mass aspect.

When $\Psi $ is expanded in spherical harmonics,

\begin{equation}
\Psi =\Psi ^{0}+\Psi ^{i}Y_{1i}^{0}+\Psi ^{ij}Y_{2ij}^{0}+...,  \label{cc}
\end{equation}%
one has, from the work of Hermann Bondi\cite{Bondi,N.T.}, the physical
meaning of the $l=(0,1)$ harmonics as the mass and the linear 3-momentum:
i.e.,

\begin{eqnarray}
\Psi ^{0} &=&-\frac{2\sqrt{2}G}{c^{2}}M,  \label{m} \\
\Psi ^{i} &=&-\frac{6G}{c^{3}}P^{i}.  \label{p}
\end{eqnarray}

The first result from the reality is that the $M$ and $P^{i}$ are real. We
will see shortly that it also leads to an expression for conservation of 
\textit{angular momentum} in the form of a flux law.

Returning to the expression for $\Psi ,$ we see that since $\sigma ^{0}$ is 2%
$^{nd}$ order the term$\sqrt{2}\sigma ^{0}(\overline{\sigma }^{0})^{\prime }$
can be ignored and the condition becomes

\begin{equation*}
\Psi =\psi _{2}^{0\,}+\text{\dh }^{2}\overline{\sigma }^{0}=\overline{\psi }%
_{2}^{0\,}+\overline{\text{\dh }}^{2}\sigma ^{0}.
\end{equation*}%
Substituting the expressions for $\psi _{2}^{0\,}$ and $\sigma ^{0}$ (i.e.,
Eqs.(\ref{12}) and (\ref{N})) into $\Psi $ we obtain, after some
cancellation of terms 
\begin{equation}
\Psi =\chi +{\Large (}\frac{\sqrt{2}}{2}a^{k\prime }-2kiD^{i\prime }\bar{D}%
^{j\prime \prime }\epsilon _{ijk}{\Large )}Y_{1k}^{0}+k\frac{\sqrt{2}}{6}%
\int (\bar{D}^{i\prime \prime }D^{j\prime \prime })du_{r}Y_{2ij}^{0}.
\label{dd}
\end{equation}

The $l=2$ harmonic is real without any added conditions. This follows from
the symmetry of $Y_{2ij}^{0}.$

The $l=0$ harmonic condition is just 
\begin{equation}
\Psi ^{0}=\chi =\overline{\chi }=-\frac{2\sqrt{2}G}{c^{2}}M,  \label{ee}
\end{equation}%
while the $l=1$ condition, with $a^{k}=a_{R}^{k}+ia_{I}^{k},$ is that 
\begin{equation}
a_{I}^{k\prime }=2\sqrt{2}k[D_{E}^{i\prime }D_{E}^{j\prime \prime
}+D_{M}^{i\prime }D_{M}^{j\prime \prime }]\epsilon _{ijk}.  \label{ff}
\end{equation}

\qquad \textbf{Comment: }\textit{In an earlier section we saw that the
complex} $a^{k}$ \textit{was determined by Eq.}.(\ref{C2}), i.e., by\textbf{%
\ }%
\begin{equation}
a^{k\prime \prime \prime }=2\sqrt{2}ki(D^{i\prime }\bar{D}^{j\prime \prime
})^{\prime \prime }\epsilon _{ijk}+\sqrt{2}ki(\bar{D}^{i\prime \prime
}D^{j\prime \prime })^{\prime }\epsilon _{ijk}.  \label{fff}
\end{equation}

\qquad \textit{It was a relief to us to see that the\ imaginary part of Eq.}(%
\ref{fff}) \textit{coincided with the reality condition, Eq}.(\ref{ff}). 
\textit{For completeness we give the real part of Eq.}(\ref{fff}):

\begin{equation}
a_{R}^{k\prime \prime \prime }=2\sqrt{2}k(D_{M}^{j\prime }D_{E}^{i\prime
})^{\prime \prime \prime }\epsilon _{ijk}-\sqrt{2}k(D_{E}^{i\prime \prime
}D_{M}^{j\prime \prime })^{\prime }\epsilon _{ijk}.  \label{gg}
\end{equation}

The linear momentum given by 
\begin{equation}
\Psi ^{i}=-\frac{6G}{c^{3}}P^{i}=[\frac{\sqrt{2}}{2}a_{R}^{k}-2k(D_{M}^{j%
\prime }D_{E}^{i\prime })\epsilon _{ijk}]^{\prime }.  \label{hh}
\end{equation}

The last piece of information that we obtain from the reality condition
comes from the conventional identification in linear theory of source
angular momentum as observed from infinity, namely: 
\begin{equation}
J^{k}=\frac{\sqrt{2}c^{3}}{12G}\psi _{1k}^{0}|_{I}.  \label{ii}
\end{equation}

From Eq.(\ref{11}) we see that

\begin{equation}
J^{k}=\frac{\sqrt{2}c^{3}}{12G}a_{I}^{k},  \label{jj}
\end{equation}%
so that, from Eq.(\ref{ff}), with the restored c's from $(^{\prime })=>$ $%
c^{-1}(^{\prime }),$ we obtain 
\begin{equation}
J^{k\prime }=\frac{2}{3c}[D_{E}^{i\prime }D_{E}^{j\prime \prime
}+D_{M}^{i\prime }D_{M}^{j\prime \prime }]\epsilon _{ijk}.  \label{kk}
\end{equation}

In other words the reality condition applied to the $l=1$ harmonic, i.e.,
Eq.(\ref{ff}), \textit{describes the loss of angular momentum}. \textit{This
agrees exactly with the classical loss of angular momentum due to
electromagnetic dipole radiation}\cite{L.L.}.

\section{The Metric}

In our Bondi coordinate system, the inverse metric takes the form\cite{N.T.}:

$\qquad \qquad \qquad \qquad \qquad \qquad g^{ab}=\left[ 
\begin{array}{cccc}
0 & 1 & 0 & 0 \\ 
1 & g^{11} & g^{1\zeta } & g^{1\overline{\zeta }} \\ 
0 & g^{1\zeta } & g^{\zeta \zeta } & g^{\zeta \overline{\zeta }} \\ 
0 & g^{1\overline{\zeta }} & g^{\zeta \overline{\zeta }} & g^{\overline{%
\zeta }\overline{\zeta }}%
\end{array}%
\right] ,$

\noindent where all the metric components are given in terms of metric
coefficients $U$, $\omega $, $X^{A}$, and $\xi ^{A}$, for $A\in \{\zeta ,%
\bar{\zeta}\}$. 
\begin{equation}
\begin{array}{l}
g^{11}=2(U-\omega \bar{\omega}), \\ 
g^{1A}=X^{A}-(\bar{\omega}\xi ^{A}+\omega \bar{\xi}^{A}), \\ 
g^{AB}=-(\xi ^{A}\bar{\xi}^{B}+\xi ^{B}\bar{\xi}^{A}),%
\end{array}
\label{20}
\end{equation}

These metric coefficients satisfy a set of radial and non-radial equations
given in terms of the spin coefficients just calculated\cite{N.T.,7}:

\begin{equation}
\begin{array}{l}
\ \frac{\partial \xi ^{A}}{\partial r}=\rho \xi ^{A}+\sigma \bar{\xi}^{A},
\\ 
\ \frac{\partial \omega }{\partial r}=\rho \omega +\sigma \bar{\omega}-(\bar{%
\alpha}+\beta ), \\ 
\ \frac{\partial X^{A}}{\partial r}=(\alpha +\bar{\beta})\xi ^{A}+(\bar{%
\alpha}+\beta )\bar{\xi}^{A}, \\ 
\ \frac{\partial U}{\partial r}=(\bar{\alpha}+\beta )\bar{\omega}+(\alpha +%
\bar{\beta})\omega -\gamma -\bar{\gamma},%
\end{array}
\label{21}
\end{equation}

\begin{equation}
\begin{array}{l}
\delta X^{A}-\Delta \xi ^{A}=(\mu +\bar{\gamma}-\gamma )\xi ^{A}+\bar{\lambda%
}\bar{\xi}^{A}, \\ 
\delta \bar{\xi}^{A}-\bar{\delta}\xi ^{A}=(\bar{\beta}-\alpha )\xi ^{A}+(%
\bar{\alpha}-\beta )\bar{\xi}^{A}, \\ 
\delta \bar{\omega}-\bar{\delta}\omega =(\bar{\beta}-\alpha )\omega +(\bar{%
\alpha}-\beta )\bar{\omega}+\mu -\bar{\mu}, \\ 
\delta U-\Delta \omega =(\mu +\bar{\gamma}-\gamma )\omega +\bar{\lambda}\bar{%
\omega}-\bar{\nu}.%
\end{array}
\label{22}
\end{equation}

Still using the flat-space derivative operators, these equations can be
integrated as was done earlier, we obtain (using certain standard coordinate
conditions for the sphere metric) the metric coefficients:

\begin{equation}
\begin{array}{cc}
\xi ^{0\zeta }=-P, & \bar{\xi}^{0\zeta }=0, \\ 
\xi ^{0\bar{\zeta}}=0, & \bar{\xi}^{0\bar{\zeta}}=-P, \\ 
P\equiv (1+\zeta \bar{\zeta}), & 
\end{array}
\label{23}
\end{equation}

\begin{equation}
\xi ^{A}=\ \frac{\xi ^{0A}}{r}-\ \frac{\bar{\xi}^{0A}\sigma ^{0}}{r^{2}},
\label{24}
\end{equation}

\begin{eqnarray}
\omega &=&-\ \frac{\bar{\eth }\sigma ^{0}}{r}-\frac{\psi _{1}^{0}}{2r^{2}}%
+ki(\frac{4D^{i}\bar{D}^{j\prime }}{3r^{3}}-\frac{\sqrt{2}D^{i}\bar{D}^{j}}{%
6r^{4}})\epsilon _{ijk}Y_{1k}^{1}  \label{25} \\
&&+k(\frac{2\sqrt{2}D^{i}\bar{D}^{j\prime }}{3r^{3}}-\frac{D^{i}\bar{D}^{j}}{%
6r^{4}})Y_{2ij}^{1}{\Large ,}  \notag
\end{eqnarray}

\begin{eqnarray}
X^{A} &=&\xi ^{0A}[\frac{\bar{\psi _{1}^{0}}}{6r^{3}}+ki(\frac{2\bar{D}%
^{i}D^{j\prime }}{3r^{4}}-\frac{\sqrt{2}\bar{D}^{i}D^{j}}{10r^{5}})\epsilon
_{ijk}Y_{1k}^{-1}  \label{26} \\
&&-k(\frac{2\sqrt{2}\bar{D}^{i}D^{j\prime }}{3r^{4}}-\frac{\bar{D}^{i}D^{j}}{%
10r^{5}})Y_{2ij}^{-1}]  \notag \\
&&+\bar{\xi}^{0A}[\frac{\psi _{1}^{0}}{6r^{3}}-ki(\frac{2D^{i}\bar{D}%
^{j\prime }}{3r^{4}}-\frac{\sqrt{2}\bar{D}^{j}D^{i}}{10r^{5}})\epsilon
_{ijk}Y_{1k}^{1}  \notag \\
&&-k(\frac{2\sqrt{2}D^{i}\bar{D}^{j\prime }}{3r^{4}}-\frac{\bar{D}^{j}D^{i}}{%
10r^{5}})Y_{2ij}^{-1}]{\Large ,}  \notag
\end{eqnarray}

\begin{equation}
U=-1-\ \frac{\psi _{2}^{0}+\bar{\psi}_{2}^{0}}{2r}+\mathcal{A},  \label{27}
\end{equation}%
where we have bundled higher powers of $r$ into the term $\mathcal{A}$ given
explicitly by:

\begin{eqnarray}
\mathcal{A} &\equiv &\ \frac{\bar{\eth }\psi _{1}^{0}+\eth \bar{\psi}_{1}^{0}%
}{6r^{2}}{\small -}\frac{k}{3}{\Large (}\frac{10D^{i\prime }\bar{D}^{j\prime
}}{3r^{2}}{\small -}\frac{(1+7\sqrt{2})(D^{i}\bar{D}^{j})^{\prime }}{9r^{3}}+%
\frac{13D^{i}\bar{D}^{j}}{15r^{4}}{\Large )}\delta _{ij}  \label{28} \\
&&-\ \frac{ki(2\bar{D}^{j\prime }D^{i}-(\bar{D}^{j}D^{i})^{\prime })}{3r^{3}}%
\epsilon _{ijk}Y_{1k}^{0}  \notag \\
&&-k{\Large (}\frac{2D^{i\prime }\bar{D}^{j\prime }}{3r^{2}}+\frac{(1-4\sqrt{%
2})(D^{i}\bar{D}^{j})^{\prime }+30\sqrt{2}D^{i}\bar{D}^{j\prime }}{108r^{3}}-%
\frac{7D^{i}\bar{D}^{j}}{90r^{4}}{\Large )}Y_{2ij}^{0}.  \notag
\end{eqnarray}

We can now compute the inverse metric to second-order in the perturbation,
where the components given in Eq.(\ref{20}) take the form: 
\begin{equation}
g^{11}=-2-\ \frac{(\psi _{2}^{0}+\bar{\psi}_{2}^{0})}{r}+2\mathcal{A},
\label{29}
\end{equation}

\begin{eqnarray}
g^{1\zeta } &=&-P{\Large [}\frac{\eth \bar{\sigma}^{0}}{r^{2}}+\ \frac{2\bar{%
\psi}_{1}^{0}}{3r^{3}}+ik(\frac{2\bar{D}^{i}D^{j\prime }}{r^{4}}-\frac{4%
\sqrt{2}\bar{D}^{i}D^{j}}{15r^{5}})\epsilon _{ijk}Y_{1k}^{-1}  \label{30} \\
&&-k(\frac{4\sqrt{2}\bar{D}^{i}D^{j\prime }}{3r^{4}}-\frac{4\bar{D}^{i}D^{j}%
}{15r^{5}})Y_{2ij}^{-1}{\Large ]},  \notag
\end{eqnarray}

\begin{eqnarray}
g^{1\bar{\zeta}} &=&-P{\Large [}\frac{\bar{\eth }\sigma ^{0}}{r^{2}}+\ \frac{%
2\psi _{1}^{0}}{3r^{3}}-\ ik(\frac{2D^{i}\bar{D}^{j\prime }}{r^{4}}-\frac{4%
\sqrt{2}\bar{D}^{j}D^{i}}{15r^{5}})\epsilon _{ijk}Y_{1k}^{1}  \label{31} \\
&&-k(\frac{4\sqrt{2}D^{i}\bar{D}^{j\prime }}{3r^{4}}-\frac{4\bar{D}^{i}D^{j}%
}{15r^{5}})Y_{2ij}^{-1}{\Large ],}  \notag
\end{eqnarray}

\begin{eqnarray}
g^{\zeta \zeta } &=&\ \frac{2\bar{\sigma}^{0}P^{2}}{r^{3}},  \label{32} \\
g^{\bar{\zeta}\bar{\zeta}} &=&\ \frac{2\sigma ^{0}P^{2}}{r^{3}},  \notag \\
g^{\zeta \bar{\zeta}} &=&\ \frac{-P^{2}}{r^{2}}.  \notag
\end{eqnarray}

Calculating the inverse of $g^{ab}.i.e.,$

$\qquad \qquad \qquad \qquad \qquad g_{ab}=\left[ 
\begin{array}{cccc}
g_{00} & 1 & g_{0\zeta } & g_{0\overline{\zeta }} \\ 
1 & 0 & 0 & 0 \\ 
g_{0\zeta } & 0 & g_{\zeta \zeta } & g_{\zeta \overline{\zeta }} \\ 
g_{0\overline{\zeta }} & 0 & g_{\zeta \overline{\zeta }} & g_{\overline{%
\zeta }\overline{\zeta }}%
\end{array}%
\right] ,$

\noindent we obtain the metric, now fully expressed in terms of Maxwell
field or in terms of the complex dipole, $D^{i}:$

\begin{eqnarray}
g_{00} &=&2+r^{-1}{\Large \{}2\chi +[\frac{\sqrt{2}}{2}(a^{k\prime }+\bar{a}%
^{k\prime })-2ki((D^{i\prime }\bar{D}^{j\prime })^{\prime })\epsilon
_{ijk}]Y_{1k}^{0}{\Large \}}  \label{33} \\
&&+r^{-1}k[\frac{\sqrt{2}}{2}(D^{i}\bar{D}^{j\prime \prime }+\bar{D}%
^{i}D^{j\prime \prime })^{\prime }+\ \ \frac{\sqrt{2}(\bar{D}^{i\prime
}D^{j\prime })^{\prime }}{3}]Y_{2ij}^{0}-2\mathcal{A},  \notag
\end{eqnarray}

\begin{eqnarray}
g_{\zeta 0} &=&-P^{-1}{\Large (}\bar{\eth }\sigma ^{0}+\ \frac{2a^{k}}{3r}%
Y_{1k}^{1}-ik(\frac{D^{i}\bar{D}^{j\prime }}{r^{2}}-\frac{4\sqrt{2}D^{i}\bar{%
D}^{j}}{15r^{3}})\epsilon _{ijk}Y_{1k}^{1}  \label{34} \\
&&+k(\frac{2D^{i}\bar{D}^{j\prime \prime }}{r}-\frac{4\sqrt{2}D^{i}\bar{D}%
^{j\prime }}{3r^{2}}+\frac{4D^{i}\bar{D}^{j}}{15r^{3}})Y_{2ij}^{1}{\Large ),}
\notag
\end{eqnarray}

\begin{eqnarray}
g_{\bar{\zeta}0} &=&-P^{-1}{\Large (\eth }\bar{\sigma}^{0}+\ \frac{2\bar{a}%
^{k}}{3r}Y_{1k}^{-1}+ik(\frac{\bar{D}^{i}D^{j\prime }}{r^{2}}-\frac{4\sqrt{2}%
\bar{D}^{i}D^{j}}{15r^{3}})\epsilon _{ijk}Y_{1k}^{-1}  \label{35} \\
&&+k(\frac{2\bar{D}^{i}D^{j\prime \prime }}{r}-\frac{4\sqrt{2}\bar{D}%
^{i}D^{j\prime }}{3r^{2}}+\frac{4D^{i}\bar{D}^{j}}{15r^{3}})Y_{2ij}^{-1}%
{\Large ),}  \notag
\end{eqnarray}

\begin{eqnarray}
g_{\zeta \zeta } &=&\ \ \frac{-2\sigma ^{0}r}{P^{2}},  \label{36} \\
g_{\bar{\zeta}\bar{\zeta}} &=&\ \ \frac{-2\bar{\sigma}^{0}r}{P^{2}},  \notag
\\
g_{\zeta \bar{\zeta}} &=&\ \frac{-r^{2}}{P^{2}}.  \notag
\end{eqnarray}

We recall that%
\begin{equation}
\sigma ^{0}=\ \ \frac{k\sqrt{2}}{2}{\Large [}\frac{1}{3}\int D^{i\prime
\prime }\bar{D}^{j\prime \prime }\,du_{r}-(\bar{D}^{i}D^{j\prime \prime
})^{\prime }-\ \ \frac{2}{3}(\bar{D}^{i\prime }D^{j\prime \prime }){\Large ]}%
Y_{2ij}^{2}.  \label{37}
\end{equation}

Eqs. (\ref{33})-(\ref{37}) describe the metric resulting from the
perturbation off flat space induced by the electric and magnetic dipoles
Maxwell field.

\section{Discussion and Conclusion}

In the present work we have extended ideas and work of Bramson. Starting
with the radiating dipole Maxwell field, we found in addition to the 2nd
order Weyl tensor, the spin coefficients and metric up to second order. We
also found the physical content of the $l=1$ spherical harmonic of the
reality condition to be the angular momentum conservation law. Thus, in our
approximation we can fully characterize, up to second order, the changes
imposed on the Minkowski space-time by the electromagnetic perturbation.

Bramson, by making the error of setting $a^{k}=0,$ missed the interesting
physics contained in $a^{k}.$ Furthermore, by not continuing on to the
spin-coefficient equations he had no means to investigate the reality
conditions and would have missed the conservation of angular momentum. One
of the most important of the spin-coefficients is the Bondi asymptotic
shear. We found that the shear (or its time derivative, the Bondi
news-function) is fully determined by the electric and magnetic dipole
moments. We also saw the Bondi mass loss equation yields\textbf{\ }the
classical energy loss from electrodynamics as well as finding the Bondi
linear momentum also determined by the Maxwell field.

An interesting observation is that in the Bondi mass loss formula, the
energy is lost only through the Maxwell dipole radiation and there is no
term involving the resultant gravitational radiation, via the Bondi
news-function. The reason for this is that the gravitational radiation is
4th order and thus not visible at our approximation. It however does suggest
that whenever there is electromagnetic radiation it will always be
accompanied by weak gravitational radiation.

This approximation scheme can be applied to any number of perturbation
calculations. In a future paper, we will apply this method to a more general
perturbation off the Schwarzschild solution. One should be able in principle
to do similar things from a Reissner-Nordstr\"{o}m or Kerr background as
well, and in future work we will investigate these scenarios.

\section{Acknowledgments}

We would like to thank Carlos Kozameh and Gilberto Silva-Ortigoza for the
many discussions that helped to clarify the ideas behind this work.

\end{document}